\documentclass[12pt]{article}
\usepackage{epsfig, amsmath, amssymb}
\usepackage{graphicx,psfrag} 
\newcommand{\be}{\begin{equation}}
\newcommand{\ee}{\end{equation}}
\newcommand{\bea}{\begin{eqnarray}}
\newcommand{\eea}{\end{eqnarray}}
\newcommand{\bA}{\begin{array}}
\newcommand{\eA}{\end{array}}
\newcommand{\bc}{\begin{center}}
\newcommand{\ec}{\end{center}}
\newcommand{\al}{\alpha}

\newcommand{\ra}{\rightarrow}
\newcommand{\del}{\partial}

\newcommand{\ie}{{\it i.e.}}
\newcommand{\eg}{{\it e.g.}}

\newcommand{\Nf}{${\cal N}{=}4$}

\begin{document}


\begin{titlepage}
\vspace{30mm}

\bc

\hfill  {\tt arXiv:0909.4731 [hep-th]} 
\\         [22mm]

{\huge Null cosmological singularities\\ [2mm]
and free strings}
\vspace{16mm}

{\large K.~Narayan} \\
\vspace{3mm}
{\small \it Chennai Mathematical Institute, \\}
{\small \it SIPCOT IT Park, Padur PO, Siruseri 603103, India.\\}

\ec
\medskip
\vspace{40mm}

\begin{abstract}
We continue exploring free strings in the background of null Kasner-like 
cosmological singularities, following arXiv:0904.4532 [hep-th]. We study 
the free string Schrodinger wavefunctional along the lines of 
arXiv:0807.1517 [hep-th]. We find the wavefunctional to be nonsingular 
in the vicinity of singularities whose Kasner exponents satisfy certain 
relations. We compare this with the description in other variables. 
We then study certain regulated versions of these singularities where 
the singular region is replaced by a substringy but nonsingular region 
and study the string spectra in these backgrounds. The string modes 
can again be solved for exactly, giving some insight into how string 
oscillator states get excited near the singularity.
\end{abstract}

\end{titlepage}

\newpage 
{\footnotesize
\begin{tableofcontents}
\end{tableofcontents}
}

\vspace{2mm}

\section{Introduction}

In this note, we continue exploring free string propagation in the 
background of null cosmological singularities, following \cite{knNullws}, 
and motivated by \cite{dmnt, adnt, adnnt, adgot}, as well as some 
earlier investigations, \eg\ \cite{horowitzsteif,bhkn,cornalbacosta,lms,
lawrence,horopol,ckr,prt,david,bbop,csv0506,li0506,blauMpp,ohta,hertoghoro,
Chu:2006pa,linwen,Turok:2007ry,crapsetal,crapsetal2,niz}.

Time-dependent generalizations of AdS/CFT where the bulk contains null 
or spacelike cosmological singularities were studied in 
\cite{dmnt,adnt,adnnt}, with a nontrivial dilaton $e^\Phi$ that vanishes 
at the location of the cosmological singularity, the curvatures behaving 
as\ $R_{MN}\sim\del_M\Phi\del_N\Phi$.  The gauge theory duals are \Nf\ \
Super Yang-Mills theories with a time-dependent gauge coupling
$g_{YM}^2=e^\Phi$, and \cite{dmnt,adnt,adnnt} describe aspects of the dual
descriptions of the bulk cosmological singularities. From the bulk
point of view, supergravity breaks down and possible resolutions of
the cosmological singularity stem from stringy effects. Indeed noting
$\al'\sim{1\over\sqrt{g_sN}}={1\over g_{YM}\sqrt{N}}$ from the usual 
AdS/CFT dictionary and extrapolating naively to these time-dependent
cases with a nontrivial dilaton, we have\ 
$\al'\sim{1\over e^{\Phi/2}\sqrt{N}}$\ indicating vanishing effective 
tension for stringy excitations, when\ $e^\Phi\ra 0$\ near the
singularity. In general, we expect that stringy effects are becoming 
important near the bulk singularity, corresponding to possible gauge 
coupling effects in the dual gauge theory. It is therefore interesting 
to understand worldsheet string effects in the vicinity of the
singularity. Owing to the technical difficulties with string
quantization in an AdS background with RR flux, we would like to look
for simpler, purely gravitational backgrounds as toy models whose
singularity structure shares some essential features with the
backgrounds in the AdS/CFT investigations.

In \cite{knNullws}, we described purely gravitational spacetimes, with
no other background fields turned on, containing null Kasner-like
cosmological singularities where tidal forces diverge. The Kasner
exponents satisfy certain algebraic relations stemming from the
supergravity equations of motion. These are related by a coordinate
transformation to anisotropic plane waves with singularities. We then
studied the free string spectra in these backgrounds, aided by the
fact that the string mode functions can be exactly solved for in these
backgrounds. Using the mode asymptotics in the near singularity
region, the worldsheet Hamiltonian in lightcone gauge can be
simplified enabling a detailed analysis of the string spectrum.

The Schrodinger wavefunctional was found to be a useful diagnostic for
the response of gauge theories to time-dependent coupling sources
\cite{adnnt}: these theories are dual to AdS cosmologies with
spacelike singularities, in particular with dilaton profiles vanishing
as\ $e^\Phi\sim t^p,\ p>0$,\ with $t$ being a time-like time
coordinate.  Among other things, \cite{adnnt} found that the quantum
mechanical wave function of the system, describing its response to the
external time dependence, in general acquires a time dependent phase
factor. For $p\geq 1$, this phase becomes wildly oscillating and
diverges as $t\ra 0^-$. As a result, the wave function of the system
(in the Schrodinger picture) does not have a well defined limit as
$t\ra 0^-$.  In contrast for $p<1$, the phase factor does not diverge
and the wave function has a well defined limit as $t\ra 0^-$.  The
energy diverges in both cases. By contrast, null singularities with
$e^\Phi\sim (x^+)^p$ appear to be better defined \cite{dmnt}, with a
well-defined ``near singularity'' wavefunction and no 
divergences\footnote{Indeed, in terms of the redefined gauge fields\
${\tilde A}_\mu=e^{-\Phi/2} A_\mu$, the gauge theory interaction terms 
become unimportant near the singularity \cite{dmnt} and the lightcone 
Hamiltonian containing simply the kinetic terms gives rise to a free 
lightcone Schrodinger equation for the gauge theory wavefunctional. 
Operators involving ${\tilde A}_\mu$ are likely to not have local bulk 
duals as argued in \cite{dmnt}, again suggesting stringy effects.}.
These findings could acquire possible modifications due to gauge
coupling renormalization effects, as discussed in \cite{adnnt}.

Motivated by this, in the present paper, we continue investigating
null singularities and the string worldsheet description, following
\cite{knNullws}. Along the lines of the Schrodinger wavefunctional
analysis \cite{adnnt} described above, we study the Schrodinger
equation for the string wavefunctions using the free string
Hamiltonian to study the response of strings to null Kasner-like
cosmological singularities, thereby gaining insights into string
propagation across these null singularities. We find that for
singularities whose Kasner exponents satisfy certain relations, the
wavefunction has a well-defined limit near the singularity, as $x^+\ra
0^-$. We compare this with the corresponding description in other
variables.  We then describe certain regulated versions of these
spacetimes where the singular region is excised by a substringy but
nonsingular region (albeit by a nonanalytic regulator), and study the
string spectra in these regulated regions. The string modes can
luckily be again solved for exactly, giving some insight into how
string oscillator states turn on near the singularity. In particular,
comparing the (instantaneous) masses of string oscillator states with
the local energy (curvature) scales in the regulated near singularity 
region, we find that a finite number of oscillator states are light 
for a finite regulator.

In sec.~2, we review some key points of \cite{knNullws}, and then 
discuss the Schrodinger wavefunctional in sec.~3. In sec.~4, we 
describe various dimensional properties of these spacetimes that 
make this description consistent with the no-scale property of 
plane wave spacetimes (that these are related to by a coordinate 
transformation). Finally in sec.~5, we discuss some regulated 
versions of these singularities and string propagation in them, 
closing with a discussion in sec.~6.

\section{Reviewing null singularities and strings}

We are interested in purely gravitational spacetime backgrounds (for 
simplicity) that have a Big-Crunch (-Bang) type of singularity at 
some value of the lightlike time coordinate $x^+$. Thus consider
\be\label{solnsfh}
ds^2=e^{f(x^+)}\left(-2dx^+dx^- + dx^idx^i\right) + e^{h(x^+)}dx^mdx^m\ ,
\ee
with two null scale factors (and all other backgrounds fields vanishing). 
It is straightforward to generalize this to multiple scale factors 
$e^{h_m(x^+)}$. 
Simple classes of null Kasner-like cosmological singularities arise 
in the vicinity of $x^+=0$ with
\be\label{absolns}
ds^2 = (x^+)^a\left(-2dx^+dx^-+dx^idx^i\right) + (x^+)^b dx^mdx^m\ ,
\qquad\ \ a>0\ ,
\ee
where $i=1,2$, $m=3,\ldots,D-2$. A solution with $a<0$ can be 
transformed to one with $a>0$ by redefining\ $y^+={1\over x^+}$. These 
are Ricci-flat solutions to the Einstein equations if\ $R_{++}=0$, \ie\
\be\label{EOMfh}
{1\over 2}(f')^2-f''+{D-4\over 4} (-2h''-(h')^2+2f'h') = 0
\ \Rightarrow\ \ a^2+2a+{D-4\over 2}(-b^2+2b+2ab) = 0\ .
\ee
The first equation, in terms of the scale factors $e^f, e^h$, shows that 
the curvature for the 4D scale factor $e^f$ is sourced by those for the 
``internal space'' scale factor $e^h$: indeed the $h$ (and more generally 
$h_m$ for multiple scale factors) are the analogs of the dilaton scalar 
in the AdS/CFT cosmological context \cite{dmnt,adnt,adnnt} where the 
corresponding equation was\ $R^{(4)}_{++}={1\over 2} (\del_+\Phi)^2$. 
That is, the kinetic terms\ $(\del_+h_m)^2$\ (and related cross-terms) 
play the role of the dilaton in driving the singular behaviour of the 
4D part of the spacetime.

For any $b\neq a$, these give solutions\ 
$2a=-2-(D-4)b\pm\sqrt{4+(D-4)(D-2)b^2}$. Requiring $a>0$ means we take 
the positive radical. Requiring unambiguous analytic continuation from 
$x^+<0$ to $x^+>0$ across the singularity means $a,b$ are even integers: 
this is more restrictive but such solutions do exist\footnote{From 
eq.(10) of \cite{knNullws}, we have\
$(a,b) = (0,2), (44,-2), (44,92), (2068,-92) \ldots$, for $D=26$ 
(bosonic string), and\ $(a,b) = (0,2),(12,-2), (12,28), (180,-28), 
(180,390),\ldots$, for $D=10$ (superstring).}.
While no curvature invariants diverge in these null backgrounds, the 
Riemann components $R_{+I+I}, I=i,m,$ are nonvanishing giving diverging 
tidal forces: analysing the deviation of null geodesic congruences, we 
find the accelerations\ $a^i, a^m\sim {1\over (x^+)^{2a+2}}$. 
We refer to \cite{knNullws} for details on the various properties of 
these spacetimes and the Kasner exponents.

We would like to study string propagation in these backgrounds.
Starting with the closed string worldsheet action\ 
$S = -\int {d\tau d\sigma\over 4\pi\al'} \sqrt{-h} h^{ab}\ 
\del_a X^\mu \del_b X^\nu g_{\mu\nu}(X)$, we use lightcone gauge 
$x^+=\tau$ and set $h_{\tau\sigma}=0$, with\ 
$E(\tau,\sigma)=\sqrt{-{h_{\sigma\sigma}\over h_{\tau\tau}}}$ , as in 
\cite{Polchinski:2001ju} (see also \cite{Metsaev:2000yf}), obtaining
\be\label{wsAction}
S = -{1\over 4\pi\al'} \int d^2\sigma\ 
\left( -E g_{IJ} \del_\tau X^I \del_\tau X^J 
+ {1\over E}\ g_{IJ} \del_\sigma X^I \del_\sigma X^J 
- 2E g_{+-} \del_\tau X^- \right)\ .
\ee
Then setting the momentum conjugate to $X^-$ to a $\tau$-independent 
constant\ $p_-={E g_{+-}\over 2\pi\al'}=-{1\over 2\pi\al'}$ by a 
$\tau$-independent $\sigma$-reparametrization invariance (not fixed by 
the gauge fixing above), we obtain\ $E=-{1\over g_{+-}}$, giving 
\be\label{action}
S = {1\over 4\pi\al'} \int d^2\sigma \left( (\del_\tau X^i)^2 
- e^{2f(\tau)} (\del_\sigma X^i)^2 + e^{h(\tau)-f(\tau)} (\del_\tau X^m)^2
- e^{h(\tau)+f(\tau)} (\del_\sigma X^m)^2 \right) ,
\ee
containing only the physical transverse degrees of freedom\ 
$X^I\equiv X^i, X^m$, of the string.
We can now calculate the Hamiltonian $H[X^-,p_-,X^I,\Pi^I]$ (using\ 
$E={2\pi\al' p_-\over g_{+-}}$ in (\ref{wsAction})), and solve for 
$X^-$ using\ $\del_\tau X^-={\del H\over\del p_-}$ .

The lightcone gauge quantization of strings in these backgrounds is 
aided by the fact that the classical string modes here can be exactly 
solved for from the worldsheet equations of motion following from 
(\ref{action}): we have the mode functions\ 
\be\label{modesolns}
f^I_n(\tau) =  \sqrt{n \tau^{d_I}} \left({c^I_{n1}} 
J_{\frac{d_I}{2a+2}}\left(\frac{n \tau^{a+1}}{a+1} \right) + {c^I_{n2}} 
Y_{\frac{d_I}{2a+2}}\left(\frac{n \tau^{a+1}}{a+1} \right) \right)\ ,
\ee
where $d_I=1,2\nu$, for $I=i,m$, respectively,\ $\nu={a+1-b\over 2}$ 
and $c^I_{n1},c^I_{n2}$ are constants.
Using the basis modes $f^I_n(\tau)e^{in\sigma}$, we can then mode expand 
the worldsheet coordinate fields $X^I(\tau,\sigma)$. Then by the usual 
procedure to simplify the Hamiltonian using the mode expansion, we obtain 
\bea\label{Hamil}
&& H = {1\over 2\al'} \left(({\dot X}^i_0)^2 + \tau^{b-a}({\dot X}^m_0)^2\right)
+ {1\over 2\al'} \sum_n |k^i_n|^2 \Biggl( ( \{a^i_n,a^i_{-n}\} + 
\{{\tilde a}^i_n,{\tilde a}^i_{-n}\} ) \left( |{\dot f}^i_n|^2 + n^2 \tau^{2a} 
|f^i_n|^2 \right) \nonumber\\
&& \qquad
-\ \{a^i_n,{\tilde a}^i_n\} \left( ({\dot f}^i_n)^2 + n^2 \tau^{2a} 
(f^i_n)^2 \right) - \{a^i_{-n},{\tilde a}^i_{-n}\} 
\left( ({\dot f}^{i*}_n)^2 + n^2 \tau^{2a} (f^{i*}_n)^2 \right) \Biggr) 
\nonumber\\
&& \qquad
+\ {1\over 2\al'} \sum_n |k^m_n|^2 \Biggl( ( \{a^m_n,a^m_{-n}\} + 
\{{\tilde a}^m_n,{\tilde a}^m_{-n}\} ) \left( \tau^{b-a} |{\dot f}^m_n|^2 
+ n^2 \tau^{b+a} |f^m_n|^2 \right) \\
&& \qquad -\ \{a^m_n,{\tilde a}^m_n\} \left( \tau^{b-a} ({\dot f}^m_n)^2 
+ n^2 \tau^{b+a} (f^m_n)^2 \right) - \{a^m_{-n},{\tilde a}^m_{-n}\} 
\left( \tau^{b-a} ({\dot f}^{m*}_n)^2 + n^2 \tau^{b+a} (f^{m*}_n)^2 \right) 
\Biggr) .\nonumber
\eea

We now introduce a cutoff null surface at\ $x^+\equiv\tau=\tau_c$, akin 
to a stretched horizon outside a black hole horizon, and will evaluate 
the various mode asymptotics on this constant null-time surface.
The mode function asymptotics show distinct behaviour for the low lying
(small oscillation number $n\lesssim n_c\sim {1\over \tau_c^{a+1}}$) and 
highly stringy (large $n\gg n_c$) modes in the near singularity region. 
From the mode functions, we find the asymptotics as $\tau=\tau_c\ra 0$,
\bea\label{modeasympt}
f^I_n \ra {\lambda^I_{n0}} + {\lambda^I_{n\tau}} \tau_c^{d_I}\ ,
\qquad\qquad\qquad\qquad && \qquad {n\tau_c^{a+1}\over (a+1)}\lesssim 1\ , 
\nonumber\\
f^I_n \sim {1\over \tau_c^{a/2}}~e^{- in\tau_c^{a+1}/l(a+1)}\ , 
\qquad f^m_n \sim  {1\over \tau_c^{b/2}}~e^{- in\tau_c^{a+1}/(a+1)}\ , 
&& \qquad\  {n\tau_c^{a+1}\over (a+1)}\gg 1\ ,
\eea
where the $\lambda^I_{n0},\lambda^I_{n\tau}$ are constant coefficients 
arising from the Bessel function expansions and involving $c^I_{n1},
c^I_{n2}$. The highly stringy modes (positive frequency in the second 
line, with $c^I_{n1}=1, c^I_{n2}=-i$) are essentially ultra-short 
wavelength relative to the cutoff length scale $\tau_c$.
Note that this implies that for\ $\tau_c^{a+1}\gtrsim 1$, the $n=1$ 
oscillator state is already ``highly stringy''. Note that for any 
nonzero if infinitesimal regularization $\tau_c$, there exist such 
highly stringy modes, of sufficiently high $n$ that are oscillatory.
Details on the near singularity string spectrum can be found in 
\cite{knNullws}.
Later (sec.4) we will revisit this being explicit about the length 
scales involved, to gain insight into how oscillator states turn on 
near the singularity.

\section{Wave functions and the Schrodinger picture}

Here we will use the Schrodinger equation to describe wave functions
for near singularity string states and gain insights into free string
propagation across the singularity\footnote{Our analysis is essentially 
from the bosonic parts of the string worldsheet theory. Since the 
worldsheet fermion terms are quadratic (with covariant derivatives) 
for these purely gravitational backgrounds, we expect that including 
them will not qualitatively change our results here.}. Our analysis of 
the general Schrodinger wavefunctional has parallels with that of the 
Schrodinger wavefunctionals for gauge theory duals of AdS cosmologies 
with spacelike singularities in \cite{adnnt}.

\subsection{Wave functions, oscillator states and the 
Schrodinger equation}

From \cite{knNullws} (sec.~3), the string worldsheet Hamiltonian in 
terms of the oscillator operators for the low-lying and highly stringy 
modes is
\bea\label{hamilbOsc}
H_< &=& \pi\al' ((p_{i0})^2 + \tau^{a-b}(p_{m0})^2)
+\ \sum_n {\pi\over 2(a+1) n^2}\Biggl( {1\over |c^i_{n0}|} \left( 
b^{i\dag}_{n\tau} b^i_{n\tau} +\ n^2 \tau^{2a} b^{i\dag}_{n0} b^i_{n0} \right)
 \nonumber\\
&& \qquad\qquad\qquad\qquad\qquad
+\ {1\over |c^m_{n0}|} \left((2\nu)^2\tau^{a-b} b^{m\dag}_{n\tau} b^m_{n\tau} 
+\ n^2 \tau^{b+a} b^{m\dag}_{n0} b^m_{n0} \right) \Biggr)\ ,\nonumber\\
&& H_> \sim\ \tau^a \sum_{I;\ n\gg n_c}\ {1\over a+1}\ 
(a^I_{-n} a^I_n + {\tilde a}^I_{-n} {\tilde a}^I_n + n)\ ,
\eea
with the $b^I_n$-oscillators defined as linear combinations\ 
$b^I_{n0}=\lambda^I_{n0}a^I_n-\lambda^{I*}_{n0} {\tilde a}^I_{-n} ,\ 
b^I_{n\tau}=\lambda^I_{n\tau}a^I_n-\lambda^{I*}_{n\tau} {\tilde a}^I_{-n}$,
of the creation-annihilation operators\ $a^I_n,{\tilde a}^I_n$, for the 
low-lying modes. This can be seen by analyzing the Hamiltonian 
(\ref{Hamil}) and using the mode asymptotics (\ref{modeasympt}).

\noindent We see that the Hamiltonian does not mix the various operators 
for different $I$-directions, and also for different oscillator levels 
$n$. Thus we can write\ $H=\sum_I H_I$, where
\bea
&& H_i^< = \pi\al' (p_{i0})^2 + \sum_{n\lesssim n_c}\ 
{\pi\over 2(a+1) |c^i_{n0}| n^2} \left( b^{i\dag}_{n\tau} b^i_{n\tau} 
+\ n^2 \tau^{2a} b^{i\dag}_{n0} b^i_{n0} \right)
\nonumber\\
&& H_m^< = \pi\al' \tau^{a-b}(p_{m0})^2 +\ \sum_{n\lesssim n_c}\ 
{\pi\over 2(a+1) |c^m_{n0}| n^2} \left((2\nu)^2\tau^{a-b} b^{m\dag}_{n\tau} 
b^m_{n\tau} +\ n^2 \tau^{b+a} b^{m\dag}_{n0} b^m_{n0} \right)\ ,\nonumber\\
&& H_I^> \sim\ \tau^a \sum_{I;\ n\gg n_c}\ {1\over a+1}\ 
(a^I_{-n} a^I_n + {\tilde a}^I_{-n} {\tilde a}^I_n + n)\ .
\eea
Each $H_I$ decouples into a contribution $H_I^<$ from the low-lying 
modes ($n\lesssim n_c\sim {1\over\tau_c^{a+1}}$) and another $H_I^>$ from 
the highly stringy modes ($n\gg n_c\sim {1\over\tau_c^{a+1}}$).

In this free string limit, the general state $|\Psi\rangle$ then 
factorizes into a product of states\ $\prod_{I,<,>} |\Psi_I^<\rangle 
|\Psi_I^>\rangle$, decoupled both in the various $I$-directions as well 
as between the low-lying and highly stringy modes.
The Schrodinger equation\ $i{\del\over\del\tau} |\Psi\rangle = H|\Psi\rangle$ 
for the general state then factorizes into a set of equations for each 
$I$-direction as
\be
i {\del\over\del\tau} |\Psi_I\rangle = H_I|\Psi_I\rangle
\ee
In the near singularity limit $\tau\ra 0$, we see that these equations 
simplify and become (keeping only the leading terms)
\bea\label{SchrodOsc}
i{\del\over\del\tau} |\Psi_i^<\rangle = H_i^<|\Psi_i^<\rangle
&\sim& \left( \pi\al' (p_{i0})^2 + \sum_{n\lesssim n_c}\ 
{\pi\over 2(a+1) |c^i_{n0}| n^2}  b^{i\dag}_{n\tau} b^i_{n\tau} \right)
|\Psi_i^<\rangle \ , \nonumber\\
i{\del\over\del\tau} |\Psi_m^<\rangle = H_m^<|\Psi_m^<\rangle
&\sim& \tau^{a-b} \left( \pi\al' (p_{m0})^2 +\ \sum_{n\lesssim n_c}\ 
{(2\nu)^2 \pi\over 2(a+1) |c^m_{n0}| n^2} b^{m\dag}_{n\tau} 
b^m_{n\tau}\right) |\Psi_m^<\rangle \ , \qquad b>0\ , \nonumber\\
&\sim& \tau^{a+b} \left( \sum_{n\lesssim n_c}\ 
{\pi\over 2(a+1) |c^m_{n0}|}  b^{m\dag}_{n0} b^m_{n0}\right) 
|\Psi_m^<\rangle\ , \qquad\qquad b<0\ , \nonumber\\
i{\del\over\del\tau} |\Psi_I^>\rangle = H_I^<|\Psi_I^<\rangle
&\sim& \tau^a \left( \sum_{I;\ n\gg n_c}\ {1\over a+1}\ 
(a^I_{-n} a^I_n + {\tilde a}^I_{-n} {\tilde a}^I_n + n)\right) 
|\Psi_I^>\rangle\ .
\eea
These equations can be recast as Schrodinger equations with 
time-independent Hamiltonians\ \
$i\del_{\lambda_I} |\Psi_I\rangle = H_I' |\Psi\rangle$ (with 
$\del_{\lambda_I}H_I=0$) in terms of some time parameter $\lambda_I$ 
depending on the particular modes in question. For instance, 
$\lambda_I^>=\int\tau^ad\tau={\tau^{a+1}\over a+1}$ for the highly 
stringy modes in the last line above. Thus we see that 
$|\Psi_i^<\rangle$ is well-defined across $\tau\ra 0$ as long this 
time-independent Schrodinger equation has a well-defined time 
parameter $\lambda_I$. From the form of these equations, we see that 
for both $b>0, b<0$, the states $|\Psi_m^<\rangle$ are well-defined 
as long as\ $\tau^{a+1-b}=\tau^{2\nu}$ is well-defined as $\tau\ra 0$, 
\ie\ if $2\nu\geq 0$.

The highly stringy modes are very high frequency and essentially do 
not see the time-dependence of the background at all, giving a free 
Schrodinger equation (in flat space effectively) in the last line 
of (\ref{SchrodOsc}). Thus the states $|\Psi_I^>\rangle$ are 
well-defined as $\tau\ra 0$.

Let us now make a comment on a point particle propagating in these 
backgrounds, with action\ 
$S=\int d\tau\ {1\over 2} \xi(\tau) g_{IJ}{\dot x}^I{\dot x}^J$, where 
$\xi(\tau)$ is the worldline metric. Fixing lightcone gauge $x^+=\tau$, 
the lightcone momentum $p_-={\xi(\tau) g_{+-}\over 2}$ is conserved. 
This gives the conjugate momenta\ $p_I=\xi(\tau) g_{II} {\dot x}^I$\ 
and the Hamiltonian\ $H = {1\over 2 \xi(\tau) g_{II}} p_I^2 
= {1\over 2p_-} ( p_i^2 + \tau^{a-b} p_m^2 )$.\ We can also solve for 
$x^-$ using\ $\del_\tau x^-={\del H\over\del p_-}$ , using\
$\xi(\tau)={2p_-\over g_{+-}}$ .
The fact that this point particle Hamiltonian appears well-defined 
dovetails with the fact that the low-lying oscillator modes of the 
string have asymptotics similar to the zero mode (point particle). 
Note however that the spacetimes in question are singular only due 
to diverging tidal forces arising in congruences of null geodesics. 
This suggests that such a singularity might reflect in wave propagation 
of field modes in the near singularity region, rather than in single 
particle propagation. It would be interesting to explore this further.

\subsection{The Schrodinger wave functional}

In \cite{knNullws} (and reviewed earlier), we described a
``nuts-and-bolts'' approach to string propagation near these null
Kasner-like singularities, by solving for the worldsheet string mode
functions, constructing the Hamiltonian and thence the near singularity 
string spectrum. This then leads to the Schrodinger equation
description of the string wavefunction in the previous subsection. Here 
we will describe a more general Schrodinger wave functional for string
states. This has parallels with the analysis of \cite{adnnt}. We will 
see how this dovetails with the earlier analysis.

The worldsheet string Hamiltonian following from the action 
(\ref{action}) is
\be\label{Hamilws}
H = {1\over 4\pi\al'} \int d\sigma \left[ (2\pi\al')^2 (\Pi^i)^2 
+ \tau^{2a} (\del_\sigma X^i)^2 + (2\pi\al')^2 \tau^{a-b} (\Pi^m)^2 
+ \tau^{a+b} (\del_\sigma X^m)^2 \right]\ .
\ee
As we will elaborate on in sec. 4 (incorporating length scales in 
this system), the range of $\int d\sigma$ is $\int_0^{2\pi|p_-|\al'} d\sigma$, 
involving the lightcone momentum $p_-$. This Hamiltonian is the 
physical Hamiltonian\ $H=-p_+$ satisfying the physical state condition\ 
$m^2=-2g^{+-}p_+p_--g^{II}(p_{I0})^2$. Let us denote by\ 
$\Psi[X^I(\sigma),\tau]$ the wavefunctional for string fields 
$X^I(\sigma)\equiv X^i(\sigma),X^m(\sigma)$. Then the Schrodinger 
equation for the wave functional and the functional momentum operator is
\be
i\del_\tau\Psi[X^I,\tau] = H[X^I,\tau] \Psi[X^I,\tau]\ ,\qquad\quad
\Pi^I[\sigma] = -i{\delta\over\delta X^I[\sigma]}\ .
\ee
In lightcone gauge $x^+=\tau$, this is essentially the Schrodinger 
equation for the evolution in spacetime $i{\del\over\del x^+} \Psi$ 
of the string wavefunctional.
Since spatial translations are symmetries on the worldsheet, spatial 
momenta are conserved. Now we see that the Hamiltonian is simply the 
sum $H=\sum_I H_I[X^I]$ of decoupled contributions from each of the 
string coordinate fields $X^I$. Therefore in the free string limit, it 
is consistent to assume that the string wavefunctional also factorizes 
into decoupled pieces as
\be
\Psi[X^I]=\prod_I \Psi_I[X^I] = \prod_{i,m} \Psi^i[X^i] \Psi^m[X^m]\ .
\ee
Then the Schrodinger equation becomes
\be
\sum_I {i\del_\tau\Psi_I\over \Psi_I} = {\sum_I H_I[X^I] \Psi\over \Psi}
= \sum_I {H_I \Psi_I\over \Psi_I}\ .
\ee
Thus it is consistent to assume that this equation is separable into 
decoupled equations for each string coordinate field as
\be
i\del_\tau\Psi_I[X^I,\tau] = H_I[X^I] \Psi^I[X^I,\tau]\ .
\ee
For the $X^i$, this becomes (since $a>0$)
\be\label{SchrodXi}
i\del_\tau\Psi_i[X^i,\tau] =  \int {d\sigma\over 4\pi\al'} \left[ 
(2\pi\al')^2 (\Pi^i)^2 + \tau^{2a} (\del_\sigma X^i)^2 \right] \Psi^i[X^i,\tau]
\ \ra^{\tau\ra 0}\  
-\pi\al' \int d\sigma {\delta^2\over \delta X^i[\sigma]^2} 
\Psi^i[X^i,\tau]\ ,
\ee
This is the Schrodinger equation for free propagation in a 
time-independent background, as in flat space: thus we conclude that 
free string propagation in the $x^i$-directions is nonsingular.

For the $X^m$, the Schrodinger equation becomes
\be
i\del_\tau\Psi_m[X^m,\tau] = {1\over 4\pi\al'} \int d\sigma \left[
- (2\pi\al')^2 \tau^{a-b} {\delta^2\over\delta X^m[\sigma]^2} 
+ \tau^{a+b}  (\del_\sigma X^m)^2 \right] \Psi^m[X^m,\tau] \ .
\ee
This has different behaviour depending on the Kasner exponents $a,b$.
For $b>0$, the kinetic term dominates as $\tau\ra 0$ and we have
\be\label{SchrodXmb>01}
i\del_\tau\Psi_m[X^m,\tau] = - \pi\al'\tau^{a-b} 
\int d\sigma {\delta^2\over\delta X^m[\sigma]^2} \Psi^m[X^m,\tau]\ ,
\ee
which can be recast as the free Schrodinger equation 
\be\label{SchrodXmb>02}
i\del_\lambda\Psi_m[X^m,\lambda] = - \pi\al'
\int d\sigma {\delta^2\over\delta X^m[\sigma]^2} \Psi^m[X^m,\lambda]\ ,
\qquad\quad  \lambda = \int d\tau~\tau^{a-b} = {\tau^{2\nu}\over 2\nu} \ .
\ee
This is well-defined for $2\nu=a+1-b\geq 0$. Alternatively, we can 
solve for the time dependence of (\ref{SchrodXmb>01}) to obtain
\be
\Psi[X^m,\tau] = e^{i\pi\al' {\tau^{2\nu}\over 2\nu} \int d\sigma 
{\delta^2\over\delta X^m[\sigma]^2} }\ \Psi[X^m]\ .
\ee
The phase in the functional operator is well-defined if $2\nu\geq 0$:
else we obtain a ``wildly'' oscillating phase as $\tau\ra 0$.

For $b<0$, the potential term dominates and the Schrodinger equation 
becomes
\bea\label{SchrodXmb<01}
&& i\del_\tau\Psi_m[X^m,\tau] = {\tau^{a+b}\over 4\pi\al'} \int d\sigma
(\del_\sigma X^m)^2 \Psi^m[X^m,\tau]\ , \qquad\qquad\qquad [b<0]\ , \\
\label{SchrodXmb<02}
\Longrightarrow && 
i\del_\lambda\Psi_m[X^m,\lambda] = {1\over 4\pi\al'} 
\int d\sigma (\del_\sigma X^m)^2 \Psi^m[X^m,\lambda]\ ,\qquad\quad
\lambda = {\tau^{a+b+1}\over a+b+1} = {\tau^{2\nu}\over 2\nu}\ ,
\eea
which is again well-defined if $2\nu=a+1-|b|\geq 0$. Alternatively, we
can solve for the time-dependent phase of the wavefunction as\ 
$\Psi[X^m,\tau] \sim e^{-i{\tau^{2\nu}\over 8\pi\nu\al'} \int d\sigma 
(\del_\sigma X^m)^2}\ \Psi[X^m]$,\ with a well-defined phase if $2\nu\geq 0$.

Now to see the highly stringy modes, we write the Hamiltonian 
(\ref{Hamilws}) as\
\be
H^>[X^I] = { \tau^a\over 4\pi\al'} \int d\sigma \left( {(2\pi\al')^2(\Pi^I)^2 
\over g_{II}} + g_{II} (\del_\sigma X^I)^2 \right) = {\tau^a\over 4\pi\al'} 
\int d\sigma \left( a^I_\sigma a^{I\dag}_\sigma + a^{I\dag}_\sigma 
a^I_\sigma \right)\ ,
\ee
in terms of the instantaneous creation-annihilation operators\ 
$a^I_\sigma = {1\over \sqrt{2}} \left( \sqrt{g_{II}} (\del_\sigma X^I) + 
{i(2\pi\al') \Pi^I\over\sqrt{g_{II}}} \right)$. This rewriting of the 
Hamiltonian above is only valid for modes that are sufficiently high 
frequency that the background time dependence appears frozen to them: 
indeed apart from the $\tau^a$ prefactor, this is essentially the flat 
space string Hamiltonian as it should be for such modes. Recasting in 
terms of mode expansions, we see that the Hamiltonian above is 
essentially the same as that for the highly stringy modes (\ie\ $H^>$, 
last line of (\ref{hamilbOsc})). The corresponding Schrodinger equation 
then is
\be\label{Schrodhighfreq}
i\del_\tau\Psi^>[X^I,\tau] = H^>[X^I] \Psi^>[X^I,\tau]\quad \Rightarrow
\quad i\del_{\tau^{a+1}/(a+1)}\Psi^>[X^I,\tau] = H^>_{flat} \Psi^>[X^I,\tau]\ .
\ee

The form of equations (\ref{SchrodXi}), (\ref{SchrodXmb>01}), 
(\ref{SchrodXmb<01}) and (\ref{Schrodhighfreq}) is similar to 
(\ref{SchrodOsc}).

Thus the general Schrodinger picture wavefunctional $\Psi[X^I(\sigma)]$ 
description recovers the earlier ``nuts-and-bolts'' description and is 
consistent with various detailed aspects of the spectrum and
wavefunctions described in \cite{knNullws} and in the previous
subsection. This suggests that string propagation is well-defined
across null singularities with $2\nu=a+1-b\geq 0$. These are in fact 
the singularities where the classical near-singularity string mode 
functions do not diverge \cite{knNullws}.

\subsection{Other variables and the Schrodinger wavefunctional}

Here we compare the above with the corresponding description in 
other variables that arise if we use other spacetime coordinates.

In terms of the affine parameter $\lambda={(x^+)^{a+1}\over a+1}$, we 
can write these spacetimes as 
\be
ds^2=-2d\lambda dx^- + \lambda^{A_I} (dx^I)^2\ ,\qquad A_I={a_I\over a+1}\ .
\ee
Now $g_{+-}=-1$ and fixing lightcone gauge $\tau=\lambda$, the string 
worldsheet Hamiltonian becomes
\be
H = {1\over 4\pi\al'} \int d\sigma\ \left( (2\pi\al')^2{(\Pi^I)^2\over g_{II}} 
+ g_{II} (\del_\sigma X^I)^2 \right) = {1\over 4\pi\al'} \int d\sigma\ 
\left( (2\pi\al')^2{(\Pi^I)^2\over\tau^{A_I}} + 
\tau^{A_I} (\del_\sigma X^I)^2 \right)\ ,
\ee
Now consider $A_I>0$. Then as $\tau\ra 0$, the kinetic terms dominate 
and the Schrodinger equation becomes 
\be\label{SEA_I>0}
i\del_\tau\Psi[X^I,\tau] = -\pi\al' \tau^{-A_I} \int d\sigma\ 
{\delta^2\over\delta {X^I}^2}\ \Psi[X^I,\tau]
\ee
giving for the time-dependence
\be
\Psi[X^I,\tau] \sim\ e^{-i\pi\al' {\tau^{1-A_I}\over 1-A_I} \int d\sigma\
{\delta^2\over\delta {X^I}^2}}\ \Psi[X^I]\ .
\ee
The phase in the functional operator is well-defined if $A_I<1$. 
Alternatively, we can recast (\ref{SEA_I>0}) as a free Schrodinger 
equation in terms of the time parameter $\tau^{1-A_I}$ which is 
well-defined if $A_I<1$.

For spacetimes with $A_I<0$, the potential terms dominate and we have
\be
i\del_\tau\Psi[X^I,\tau] = {\tau^{-|A_I|}\over 4\pi\al'} \int d\sigma\ 
(\del_\sigma X^I)^2 \ \Psi[X^I,\tau]\ ,
\ee
giving for the time-dependence
\be
\Psi[X^I,\tau] \sim\ e^{-i {\tau^{1-|A_I|}\over 4\pi\al'(1-|A_I|)} 
\int d\sigma\ (\del_\sigma X^I)^2} \ \Psi[X^I]\ ,
\ee
which is well-defined for $|A_I|<1$.

The condition $|A_I|<1$ is equivalent to $2\nu\geq 0$ (\eg\ ${b\over a+1}
\leq 1$) stated earlier. These spacetimes are in a sense the analogs of 
the cases $p<1$ with a well-defined wavefunctional phase arising in 
the gauge theories dual to AdS cosmologies with spacelike singularities 
\cite{adnnt}. There is no manifest time-dependent divergence in the 
Hamiltonian here, unlike the dilaton prefactor there.

Now we discuss Brinkman coordinates. From \cite{knNullws}, we know that 
the coordinate transformation\ $x^I=(x^+)^{-a_I/2} y^I$,\ where\ 
$a_I\equiv a,b$,\ recasts the spacetimes (\ref{absolns}) in manifest 
plane-wave form (this is also valid for singularities with multiple 
Kasner exponents)
\be\label{planewave}
ds^2 = -2(x^+)^a dx^+dy^- + 
\left[\sum_I \left({a_I^2\over 4}-{a_I(a+1)\over 2}\right) (y^I)^2\right] 
{(dx^+)^2\over (x^+)^2} + (dy^I)^2\ .
\ee
Here we have redefined\ $y^-=x^-+({\sum_Ia_I(y^I)^2\over 4(x^+)^{a+1}})$. 
For $a_I=a,b$ distinct, these are in general anisotropic plane-waves 
with singularities. After further redefining to the affine parameter\ 
$\lambda={(x^+)^{a+1}\over a+1}$, we obtain the metric
\be\label{planewave2}
ds^2=-2d\lambda dy^-+\sum_I\chi_I(y^I)^2{d\lambda^2\over\lambda^2}+(dy^I)^2\ ,
\qquad \chi_I={A_I\over 4} (A_I-2)\ .
\ee
The string worldsheet Hamiltonian in lightcone gauge $\tau=\lambda$ is\ 
\be
H={1\over 4\pi\al'} \int d\sigma\ \left( (2\pi\al')^2(\Pi_y^I)^2 +
(\del_\sigma y^I)^2 -\sum_I{\chi_I\over\tau^2}(y^I)^2 \right) .
\ee
We see that the Hamiltonian in these variables $y^I$ contains a 
mass-term which diverges as\ $\tau\ra 0$. The wavefunctional then 
acquires a ``wildly'' oscillating phase as $\tau\ra 0$
\be
\Psi[y^I,\tau] \sim\ e^{-{i\over\tau} \sum_I\chi_I(y^I)^2}\ \Psi[y^I]\ .
\ee
This renders a well-defined Schrodinger wavefunctional interpretation 
near the singularity difficult in these Brinkman coordinates.\\
In the Rosen-like coordinates with the variables $X^I$, as we have 
seen, there is no such divergent mass-term and the Schrodinger 
wavefunctional shows smooth behaviour across the singularity for 
spacetimes with $|A_I|<1$. In some sense, this is akin to the 
difference between the $X$ and ${\tilde X}$ dual gauge theory variables 
discussed in \cite{adnnt}. Of course physical observables defined 
appropriately presumably are well-defined independent of the choice 
of variables, although they might be more transparent in some variables.

\section{Null singularities, strings and length scales}

We now describe some relevant length scales that arise in string 
propagation in the vicinity of null cosmological singularities, 
essentially drawing various results from \cite{knNullws} but being 
explicit about length scales. Our goal is to gain insights into how 
string oscillator states get excited in the near singularity region.
In the next section, we will study regulated versions of the 
singularity which will render further support to this picture.

The no-scale nature of these spacetimes, \ie\ requiring no explicit 
length scale, is manifest in the plane-wave (Brinkman) form 
(\ref{planewave}) (\ref{planewave2}). In the Rosen coordinates 
(\ref{absolns}), where the null cosmology interpretation is manifest, 
the dimensions of various coordinates are nontrivial, as they should 
be to maintain the no-scale property. In particular, the affine 
parameter $\lambda={(x^+)^{a+1}\over a+1}$ is of dimension length 
($L$), so that $dim~x^+\equiv L^{1/(a+1)}$. This implies that\ $dim~x^i
\equiv L^{1-a/(2(a+1))}$ and\ $dim~x^m\equiv L^{1-b/(2(a+1))}$. The length 
scale characterizing the near singularity region is set by the tidal 
forces, or the acceleration: this gives the scale of curvature as\ 
$a^i\equiv M_c={1\over (x^+)^{2a+2}}$ of dimension ${1\over L^2}$ .

The lightcone gauge condition fixes\ 
$dim~\tau^{a+1}=dim~\sigma=L$. We now introduce the coordinate 
length $l$ of the string, so that $\int d\sigma\equiv\int_0^{2\pi l}d\sigma$.\ 
From the worldsheet Lagrangian, we then see that the momentum conjugate 
to $X^-$ is\ $p_-=-{l\over 2\pi\al'}$, so that the coordinate length $l$ 
is related to the lightcone momentum of the string as
\be
l=2\pi |p_-|\al' \equiv 2\pi p_-\al'\ .
\ee
We see that $p_-\leq 0$ and will denote $|p_-|$ by $p_-$ for convenience 
in what follows (note that $p^+=-g^{+-}p_-$ is positive but time-dependent). 
Our conventions agree with those of \cite{joetext} for flat space.
The corresponding Hamiltonian, $-p_+$, re-expressing the momenta $\Pi^I$ 
in terms of $\del_\tau X^I$, is
\be\label{Hamiltonian}
H = {1\over 4\pi\al'} \int d\sigma \left[ (\del_\tau X^i)^2 
+ e^{2f(\tau)} (\del_\sigma X^i)^2 + e^{h(\tau)-f(\tau)} (\del_\tau X^m)^2 
+ e^{h(\tau)+f(\tau)} (\del_\sigma X^m)^2 \right]
\ee
then shows that the dimensions of each term are consistent with\ 
$dim X^i=L^{1-a/(2(a+1)} ,\\ dim X^m=L^{1-b/(2(a+1)}$. 
The Hamiltonian then has\ $dim~H = dim~{1\over\tau}=L^{-1/(a+1)}$.

The mode function asymptotics for the low lying and highly stringy modes in 
the near singularity region on a cutoff null surface\ $x^+=\tau=\tau_c$ 
are
\bea\label{modeasymptl}
f^i_n \ra {\lambda^i_{n0}} + {\lambda^i_{n\tau}} {\tau_c\over l^{1/(a+1)}}\ , 
\qquad 
f^m_n \ra {\lambda^m_{n0}} + {\lambda^m_{n\tau}} 
{\tau_c^{2\nu}\over l^{2\nu/(a+1)}}\ ,\qquad {n\tau_c^{a+1}\over l(a+1)}\ll 1\ ,
\nonumber\\
f^i_n \sim {l^{a/(2(a+1))}\over \tau_c^{a/2}}~e^{- in\tau_c^{a+1}/l(a+1)}\ , 
\qquad 
f^m_n \sim  {l^{b/(2(a+1))}\over \tau_c^{b/2}}~e^{- in\tau_c^{a+1}/l(a+1)}\ , 
\qquad\  {n\tau_c^{a+1}\over l(a+1)}\gg 1\ ,
\eea
where the $\lambda^I_{n0},\lambda^I_{n\tau}$ are constant coefficients 
arising from the Bessel function expansions (and in the second line, 
we have chosen positive frequency modes, $c^I_{n1}=1, c^I_{n2}=-i$). A 
mode is highly stringy on this cutoff surface if
\be
n \gg {l\over \tau_c^{a+1}} \ \sim\  {p_-\al'\over \tau_c^{a+1}}\ ,
\ee
the characteristic scale being a combination of the lightcone momentum 
and the string scale. Thus this implies that for\ 
$\tau_c^{a+1}\gtrsim p_-\al'$, the $n=1$ oscillator state is already 
``highly stringy''.

The masses of the highly stringy states is\ 
$m^2\sim {1\over\al'} (a^{i\dag}_na^i_n +\ldots)$. Thus a single oscillator 
excitation has mass\ $m^2={n\over\al'}$. Comparing with the typical 
curvature scale (set by the tidal forces), we find
\be
{m^2\over M_c} \equiv\ {m^2\over a^i}\ \sim\ 
{n\over \al' ({1\over\tau_c^{2a+2}})}\ .
\ee
Thus oscillator states satisfying
\be
{p_-\al'\over \tau_c^{a+1}} \ll n \ll {\al'\over \tau_c^{2a+2}}
\ee
are light relative to the typical energy scales in the near singularity 
region. The first inequality is from our definition of highly stringy 
modes (second line of (\ref{modeasymptl})). This also implicitly 
requires that\ $p_-\tau_c^{a+1}\ll 1$, as $\tau\ra 0$. These are in 
a sense the ``instantaneous'' masses of string states on the surface 
$\tau=\tau_c$. This picture 
suggests that the typical tidal forces in the near singularity region 
are sufficiently high that they excite several highly stringy states. 
However for a given cutoff surface $\tau=\tau_c$ away from the 
singularity, only oscillators with $n\lesssim {\al'\over\tau_c^{2a+2}}$
are light. As $\tau_c\ra 0$, this upper cutoff on the oscillator number 
also increases. Thus as we approach the singularity, all oscillator 
states become light and get excited.

In what follows, we will study certain regulated versions of these 
singularities and string propagation in them, which vindicates the 
picture above.

\section{Regulating the singularity}

We now describe some regulated versions of such null singularities. These 
will in fact require an explicit length scale at which the singularity is 
regulated, so that they are not ``no-scale'' anymore. Some other 
regularized versions of plane-wave singularities have been discussed in 
\eg\ \cite{crapsetal2} (see also \cite{lms,fabmcgr} for interesting 
related discussions in other kinds of null singularities).

We see that some natural analytic regulators appear to violate
some energy conditions so that they are not allowed regularizations. For 
instance, consider modifying (\ref{absolns}) with a metric ansatz of the 
form (\ref{solnsfh}) where\ the 4D scale factor is now\ 
$e^f=L^a [({x^+\over L})^2+\epsilon^2]^{a/2}$.\ Thus the scale factor 
departs from the earlier one at a length scale given by $L$ (of $dim\ 
\tau$), within which the spacetime is not singular. $\epsilon$ is a small 
regulating parameter. This gives the Ricci curvature for the 4D part 
of the spacetime as\ 
\be
e^f=L^a (({x^+\over L})^2+\epsilon^2)^{a/2}\quad \Rightarrow\quad 
R_{++}^{(4)}={1\over 2} (f')^2-f''\ \ \longrightarrow^{x^+\ra 0}\ \
\ -{a\over (L\epsilon)^2} < 0\ .
\ee
Since this is essentially the 4D local energy density $T_{++}$, such a 
regulator violates energy positivity in the regulated region. Similar 
observations also hold for an analytic regulator of the form\ 
$e^f=L^a(({x^+\over L})^a+\epsilon^2)$.\\
In terms of the D-dim system, we find no solution to $R_{++}^{(D)}=0$ 
whose 4D scale factor $e^f$ is of the above form: the additional terms 
in $R_{++}^{(D)}$ in the regulated region ($x^+\ll L\epsilon$) are of 
the form\ ${-(2a+(D-4)b)\over (L\epsilon)^2}$ , which is negative 
definite\footnote{From eq.(9) of \cite{knNullws}, we have\ 
$2a+(D-4)b=-2\pm \sqrt{2+(D-4)(D-2)b^2}$. Requiring $a>0$ means we 
take the positive radical. It can then be shown that $2a+(D-4)b>0$ 
if $a,b\neq 0$.}.

This kind of a regulator can be thought of as a universal near singularity\ 
$x^+\ra 0$\ limit of\ $e^f=L^a [1-(1-\epsilon)e^{-({x^+\over L})^2}]^{a/2} 
\sim\ L^a [\epsilon+(1-\epsilon)({x^+\over L})^2]^{a/2} $, or other 
regulators, and so such an energy condition violation is a fairly basic 
problem of low energy regulators of the singularity. A similar feature 
also occurs in the AdS dilatonic null cosmologies discussed in \cite{dmnt}.

An alternative regularization, although not analytic, is 
\be\label{regulated}
e^f = L^a ({|x^+|\over L}+\epsilon)^a\ , \qquad 
e^h = L^b ({|x^+|\over L}+\epsilon)^b\ .
\ee
This does not have the problem above: we find 
\bea
&& \qquad\qquad\qquad\qquad\qquad\qquad 
R_{++}^{(4)} = {a(a+2)\over 2 (|x^+|+L\epsilon)^2}\ , 
\nonumber\\
&& R_{++}^{(D)}={1\over 2}(f')^2-f''+{D-4\over 4}(-2h''-(h')^2+2f'h')
= {a^2+2a+{D-4\over 2}(-b^2+2b+2ab) \over (|x^+|+L\epsilon)^2}\ .
\qquad
\eea
Thus this regulated system is automatically a solution to $R_{++}^{(D)}=0$
since the expression in the numerator vanishes for the original singular 
solution. The Riemann curvature components and the tidal forces, given 
by the geodesic deviation, are
\bea
R_{+i+i}={a(a+2)\over 4}\  (|x^+|+L\epsilon)^{a-2}\ , && \ 
R_{+m+m}={b(2a+2-b)\over 4}\  (|x^+|+L\epsilon)^{b-2}\ ,
\nonumber\\
a^i,\ a^m &\sim& {1\over L^{2a+2} ({|x^+|\over L}+\epsilon)^{2a+2}}\ ,
\eea
so that in the regulated region $|x^+|\ll L\epsilon$, the curvature scale 
is ${1\over (L\epsilon)^{a+1}}$, large but finite, and so are the tidal 
forces. Null geodesics propagating solely along $x^+$ (at constant 
$x^-,x^i,x^m$) with cross-section along the $x^i$ or $x^m$ directions 
have an affine parameter
\be
\lambda=const.\ \int dx^+ (|x^+|+L\epsilon)^a = 
const.\ {(|x^+|+L\epsilon)^{a+1}\over a+1}\ .
\ee
It is worth mentioning that although the regulated spacetime appears 
non-analytic, the geodesics, affine parameter and curvature are 
continuous in the regulated region as we cross $x^+=0$. 

This regulated spacetime can also be recast as a plane-wave
\be
ds^2 = -2(|x^+|+\epsilon)^a dx^+dy^- + 
\left[\sum_I \left({a_I^2\over 4}-{a_I(a+1)\over 2}\right) (y^I)^2\right] 
{(dx^+)^2\over (|x^+|+\epsilon)^2} + (dy^I)^2\ ,
\ee
whose singularity is now regulated.

There is of course nothing sacrosanct in such a regularization of the
singularity. Our purpose here is to simply use the regularization
(\ref{regulated}) as a crutch to gain insights into string oscillator
states turning on. It would be interesting to explore these further
with perhaps an analytic regulator, possibly with other fields (\eg\
the dilaton) turned on.

In the next subsection, we will find that the string spectrum can be 
solved exactly in these regulated backgrounds too.

\subsection{Strings near the regulated singularity}

We are primarily interested in the approach from early times to the 
almost-singular region to see how string oscillator states turn on, so 
the non-analyticity in the metric across $x^+=0$ will not concern us. 
Let us therefore study the spacetime (\ref{solnsfh}) with the scale 
factors (\ref{regulated}) for $\tau=x^+<0$. For simplicity, we will 
abuse notation and use $\tau=x^+$ rather than $-\tau=-x^+$. Also in 
what follows, we will denote $p_-$ to mean $|p_-|$ as before, for 
convenience.

The worldsheet action is given in (\ref{wsAction}), which we would like 
to quantize using lightcone gauge, as in \cite{knNullws}. Keeping the 
string length factors explicit, we set the momentum conjugate to $X^-$ 
to a $\tau$-independent constant\ 
$p_-={E g_{+-}l\over 2\pi\al'}=-{l\over 2\pi\al'}$ by a $\tau$-independent 
reparametrization invariance, thus obtaining\ $E=-{1\over g_{+-}}$. This 
then gives the reduced action in the second line of (\ref{wsAction}), 
containing only the physical transverse oscillation modes\ 
$X^I\equiv X^i, X^m$, of the string.

The string worldsheet equations of motion in the regulated near 
singularity region are
\bea\label{EOMreg}
\del_\tau^2 X^i - L^{2a} ({\tau\over L}+\epsilon)^{2a} \del_\sigma^2 X^i = 0\ , 
\qquad\qquad \nonumber\\
\del_\tau^2 X^m  + {b-a\over L({\tau\over L}+\epsilon)}\ \del_\tau X^m -
L^{2a} ({\tau\over L}+\epsilon)^{2a} \del_\sigma^2 X^m = 0\ .
\eea
Defining a new (dimensionless) variable $\tau'={\tau\over L}+\epsilon$, 
these can be recast as the equations of motion in the singular spacetime 
\cite{knNullws} in terms of the variable $\tau'$. Then we can read off 
the solutions for the mode functions (with\ $\nu={a+1-b\over 2}$),
\be\label{modesolnsReg}
f^I_n(\tau)=\sqrt{{nL^{d_I}\over l^{d_I/(a+1)}} 
({\tau\over L}+\epsilon)^{d_I}} \left[{c^I_{n1}} 
J_{\frac{d_I}{2a+2}}\left({nL^{a+1}({\tau\over L}+\epsilon)^{a+1}\over l(a+1)} 
\right) + {c^I_{n2}} 
 Y_{\frac{d_I}{2a+2}}\left({nL^{a+1}({\tau\over L}+\epsilon)^{a+1}\over l(a+1)} 
\right) \right] ,
\ee
where we have introduced factors of $l$ to make $f^I_n$ dimensionless. 
It is straightforward to see that removing the regulator as\ 
$\epsilon\ra 0$ reduces these mode functions to the ones in 
(\ref{modesolns}): in particular, the scale $L$ disappears as the no-scale 
singular spacetime is recovered for $\epsilon\ra 0$.

The mode expansion for the string worldsheet fields is
\be\label{modeexpXIn}
X^I(\tau,\sigma) = X^I_0(\tau) + \sum_{n=1}^\infty \left( 
k_n^I f^I_n(\tau) (a^I_n e^{in\sigma/l} + {\tilde a}^I_n e^{-in\sigma/l}) + 
k_n^{I*} f^{I*}_n(\tau) (a^I_{-n} e^{-in\sigma/l} + {\tilde a}^I_{-n} 
e^{in\sigma/l}) \right)\ .
\ee
Working out the momenta and commutation relations, it can be shown that \\
$k_n^i={i\over n} \sqrt{{\pi\al' l^{-1+1/(a+1)}\over 2|c^i_{n0}| (a+1)}} ,\ 
k_n^m={i\over n} \sqrt{{\pi\al' l^{-1+2\nu/(a+1)}\over 2|c^m_{n0}| (a+1)}}$.\ 
\ The $\al', l$ dependences can be also fixed by dimensional analysis.
The oscillator algebras are\ \
$[a^I_n,a^J_{-m}]=[{\tilde a}^I_n,{\tilde a}^J_{-m}]=n\delta^{IJ}\delta_{nm}$.\\
\noindent The Hamiltonian (\ref{Hamiltonian}) in this case simplifies to
\bea\label{HamilReg}
&& H = {l\over 2\al'} \left(({\dot X}^i_0)^2 + \tau^{b-a}({\dot X}^m_0)^2\right)
\nonumber\\
&& \quad +\ {l\over 2\al'} \sum_n |k^i_n|^2 \Biggl[ ( \{a^i_n,a^i_{-n}\} + 
\{{\tilde a}^i_n,{\tilde a}^i_{-n}\} ) \left( |{\dot f}^i_n|^2 + 
{n^2L^{2a}\over l^2} \Big({|\tau|\over L}+\epsilon\Big)^{2a} |f^i_n|^2 \right)
\nonumber\\
&& \ \ - \{a^i_n,{\tilde a}^i_n\} \left( ({\dot f}^i_n)^2 + 
{n^2L^{2a}\over l^2} \Big({|\tau|\over L}+\epsilon\Big)^{2a} (f^i_n)^2 \right) 
- \{a^i_{-n},{\tilde a}^i_{-n}\} \left( ({\dot f}^{i*}_n)^2 + 
{n^2L^{2a}\over l^2} \Big({|\tau|\over L}+\epsilon\Big)^{2a} 
(f^{i*}_n)^2 \right) \Biggr] \nonumber\\
&& \quad +\ {l\over 2\al'} \sum_n |k^m_n|^2 \Biggl[ ( \{a^m_n,a^m_{-n}\} + 
\{{\tilde a}^m_n,{\tilde a}^m_{-n}\} ) \left( \tau^{b-a} |{\dot f}^m_n|^2 
+ {n^2L^{b+a}\over l^2} \Big({|\tau|\over L}+\epsilon\Big)^{b+a} |f^m_n|^2 
\right) \nonumber\\
&& \quad\qquad -\ \{a^m_n,{\tilde a}^m_n\} \left( \tau^{b-a} ({\dot f}^m_n)^2 
+ {n^2L^{b+a}\over l^2} \Big({|\tau|\over L}+\epsilon\Big)^{b+a} (f^m_n)^2 
\right) \nonumber\\
&& \quad\qquad -\ \{a^m_{-n},{\tilde a}^m_{-n}\} 
\left( \tau^{b-a} ({\dot f}^{m*}_n)^2 + {n^2L^{b+a}\over l^2} 
\Big({|\tau|\over L}+\epsilon\Big)^{b+a} (f^{m*}_n)^2 \right) \Biggr]\ .
\eea

We can now study the asymptotics of these mode functions and thence of 
stringy states in the regulated (but highly curved) near singularity region.
In particular, focussing on the regulated region\ $\tau\ll L$,\ the 
mode functions above become
\be\label{modesolnsRegAsymp}
f^I_n(\tau)\sim\ \sqrt{{n(L\epsilon)^{d_I}\over l^{d_I/(a+1)}}}
\left[{c^I_{n1}} J_{\frac{d_I}{2a+2}}\left({n(L\epsilon)^{a+1}\over 
l(a+1)} \right) + {c^I_{n2}} 
Y_{\frac{d_I}{2a+2}}\left({n(L\epsilon)^{a+1}\over l(a+1)} \right) \right] ,
\ee
Now the mode asymptotics change depending on the cutoff length scale\ 
$L_c=(L\epsilon)^{a+1}$.

The low lying string modes (small oscillation number $n\ll {l\over L_c}$) 
have mode function asymptotics as $\tau\ra 0$
\be
f^i_n \ra {\lambda^i_{n0}} + {\lambda^i_{n\tau}} {\tau+L\epsilon\over 
l^{1/(a+1)}}\ , \qquad 
f^m_n \ra {\lambda^m_{n0}} + {\lambda^m_{n\tau}} 
{(\tau+L\epsilon)^{2\nu}\over l^{2\nu/(a+1)}}\ ,
\ee
so that\footnote{The constant coefficients $\lambda^I$, from the Bessel 
expansions, are
\bea\label{c12lambda}
{\lambda^i_{n\tau}} = \sqrt{n} \left({n\over 2a+2}\right)^{{1\over 2a+2}}\ 
{{c^i_{n1}}+{c^i_{n2}}\cot {\pi\over 2a+2} \over \Gamma({2a+3\over 2a+2})}\ ,
\qquad
{\lambda^i_{n0}} 
= -{c^i_{n2}} \sqrt{n} \left({n\over 2a+2}\right)^{-{1\over 2a+2}}\ 
{{\rm cosec} {\pi\over 2a+2}\over \Gamma({2a+1\over 2a+2})}\ ,\nonumber\\
{\lambda^m_{n\tau}} = \sqrt{n} \left({n\over 2a+2}\right)^{{\nu\over a+1}}\ 
{{c^m_{n1}}+{c^m_{n2}}\cot {\nu\pi\over a+1}\over\Gamma({a+\nu+1\over a+1})}\ ,
\qquad
{\lambda^m_{n0}} 
= -{c^m_{n2}} \sqrt{n} \left({n\over 2a+2}\right)^{-{\nu\over a+1}}\ 
{{\rm cosec} {\nu\pi\over a+1}\over \Gamma({a+1-\nu\over a+1})}\ .\quad 
\nonumber
\eea
} (for $2\nu\geq 0$)
\be
f^i_n\ra {\lambda^i_{n0}}\ , \qquad {\dot f}^i_n\ra 
{{\lambda^i_{n\tau}}\over l^{1/(a+1)}}\ , \qquad 
f^m_n\ra {\lambda^m_{n0}}\ , \qquad 
{\dot f}^m_n\ra {{\lambda^m_{n\tau}}\over l^{2\nu/(a+1)}}\ (2\nu)\ 
(\tau+L\epsilon)^{2\nu-1}\ .
\ee
Then the Hamiltonian (\ref{HamilReg}) for these low lying modes simplifies 
and can be rewritten as
\bea\label{Hamil1}
H_{<} &=& {\pi\al'\over l} ((p_{i0})^2 + \tau^{a-b}(p_{m0})^2)
+\ \sum_n {\pi\over 2(a+1) n^2}\Biggl( {1\over |c^i_{n0}|} \left( 
{1\over l^{1/(a+1)}} b^{i\dag}_{n\tau} b^i_{n\tau} +\ 
n^2 { (\tau+L\epsilon)^{2a}\over l^{2-1/(a+1)}} b^{i\dag}_{n0} b^i_{n0}\right)
\nonumber\\  && \qquad\qquad\qquad +\ {1\over |c^m_{n0}|} 
\left((2\nu)^2 {(\tau+L\epsilon)^{a-b}\over l^{2\nu/(a+1)}} b^{m\dag}_{n\tau} 
b^m_{n\tau} +\ n^2 {(\tau+L\epsilon)^{b+a}\over l^{(2\nu+2b)/(a+1)}} 
b^{m\dag}_{n0} b^m_{n0} \right) \Biggr)\ ,
\eea
where we have defined new oscillator modes (and their Hermitian conjugates)
\be\label{bmodes}
b^I_{n0} = \lambda^I_{n0} a^I_n - \lambda^{I*}_{n0}\ {\tilde a}^I_{-n}\ , 
\qquad  b^I_{n\tau} 
= \lambda^I_{n\tau} a^I_n - \lambda^{I*}_{n\tau}\ {\tilde a}^I_{-n}\ ,
\qquad I=i,m\ .
\ee
The algebra and other properties of these $b^I$ operators are as 
discussed in \cite{knNullws}. 
The string oscillator masses\ $m^2 = - 2 g^{+-} p_+ p_- - g^{II} (p_{I0})^2$ 
recalling that $p_-=-{l\over 2\pi\al'} ,\ -p_+=H$, then work out in the 
regulated region to\ (for\ $2\nu\geq 0$)
\be\label{masses}
m^2(\tau) \ra {1\over 2\al'(a+1)} \sum_{i,m;\ n\lesssim {l\over L_c}} 
\Bigl( {l^{a/(a+1)}\over(L\epsilon)^a} {N^i_{n\tau} \over n^2|c^i_{n0}|}
+ {(L\epsilon)^a\over l^{a/(a+1)}} {N^i_{n0} \over |c^i_{n0}|} 
+ {(2\nu)^2 l^{b/(a+1)}\over(L\epsilon)^b} {N^m_{n\tau}\over 
n^2|c^m_{n0}|} + {(L\epsilon)^b\over l^{b/(a+1)}} {N^m_{n0}\over |c^m_{n0}|} 
\Bigr)\ ,
\ee
defining
\be
N^i_{n\tau} = b^{i\dag}_{n\tau} b^i_{n\tau}\ ,\quad
N^i_{n0} = b^{i\dag}_{n0} b^i_{n0}\ , \qquad
N^m_{n\tau} = b^{m\dag}_{n\tau} b^m_{n\tau}\ ,
\quad N^m_{n0} = b^{m\dag}_{n0} b^m_{n0}\ .
\ee
The time-dependence in the masses shows that single excitations are light 
relative to the local curvature scale in the regulated region if
(from the $N^i_{n\tau}$ prefactor)
\be
{l^{a/(a+1)}/(\al'(L\epsilon)^a)\over 1/(L\epsilon)^{2a+2}} 
\ll 1 \ \Rightarrow\ \ 
(L\epsilon)^{a+2} \ll {\al'^{1/(a+1)}\over p_-^{a/(a+1)}}\ .
\ee
To obtain some intuition for this, note that for $a\sim 0$ (almost 
flat space), we have $L_c^2=(L\epsilon)^2\ll {\al'\over p_-^a}\sim \al'$, 
\ie\ the regulating length scale $L_c$ is substringy. Similar 
expressions can be obtained from the $N^I_{n0,\tau}$ prefactors.

For the case $2\nu<0$, the mode function asymptotics are different: in 
particular, the mode functions $f^I_n(\tau)$ grow large but are still 
finite due to the regulator. Then we can again calculate the Hamiltonian 
and the oscillator masses for this case.

Now we turn to the other asymptotic region of the modes: the modes are 
oscillatory for
\be
n \gg {l\over L_c} = {p_-\al'\over L_c}\ .
\ee
The mode function asymptotics for (these oscillatory) highly stringy modes 
in the regulated region are (for positive frequency modes with\ $c_1=1, 
c_2=-i$)
\be
f^i_n \sim {l^{a/(2(a+1))}\over 
(L\epsilon)^{a/2}}~e^{- in(L\epsilon)^{a+1}/l(a+1)}\ , 
\qquad f^m_n \sim {l^{b/(2(a+1))}\over 
(L\epsilon)^{b/2}}~e^{- in(L\epsilon)^{a+1}/l(a+1)}\ ,
\ee
and their derivatives are
\be\label{fdotlargen}
{\dot f}^i_n \sim\ \left( -{in (L\epsilon)^a\over l}-{a\over 2L\epsilon} 
\right) {e^{-in(L\epsilon)^{a+1}/(a+1)}\over (L\epsilon)^{a/2}}\ , \qquad
{\dot f}^m_n \sim\ \left( -{in(L\epsilon)^a\over l}-{b\over 2L\epsilon} 
\right) {e^{-in(L\epsilon)^{a+1}/(a+1)}\over (L\epsilon)^{b/2}}\ .
\ee

The Hamiltonian (\ref{Hamil}) for highly stringy modes then simplifies to
\be
H_{>} \sim\ {(L\epsilon)^a\over l} 
(a^I_{-n}a^I_n + {\tilde a}^I_{-n}{\tilde a}^I_n + n)\ ,
\ee
where the constant prefactor arises as\
$ {l\over 2\al'}\ l^{-1+{1\over a+1}} \al' {(L\epsilon)^a\over l^2} 
l^{{a\over a+1}}$. 
Thus the (instantaneous) masses of the highly stringy states in the 
regulated region are
\be
m^2\sim\ -g^{+-} H p_- \sim {1\over (L\epsilon)^a} {(L\epsilon)^a\over l} 
{l\over\al'}\ (N^I_n+{\tilde N}^I_n+n) 
\sim {1\over\al'} (N^I_n+{\tilde N}^I_n+n)\ . 
\ee
Relative to the local curvature scale given by\ ${1\over L_c^2}$ ,
these modes are light for oscillator states satisfying
\be\label{lightosc}
{p_-\al'\over L_c}\ll n\ll {\al'\over L_c^2}\ .
\ee
This implicitly requires\ $p_- \ll {1\over L_c}$. The number of such 
oscillator levels from is\ ${\al'\over L_c^2} (1-p_-L_c)$.
Thus for any finite\ $p_-\ll {1\over L_c}$,\ only a finite set of the 
highly stringy oscillator states are excited in the regulated near
singularity region, as expected. In the singular limit $L_c\ra 0$, all 
oscillator states are light and the number of excited oscillator states 
diverges. Conversely in the sector\ $p_-\sim {1\over L_c}$, the window 
of light highly stringy states pinches off.

As the lightcone momentum $p_-$ increases, the lower cutoff in
(\ref{lightosc}) increases and the oscillator states that are highly
stringy must have higher $n$.  Conversely, in the $p_-=0$ zero mode
sector, essentially all oscillator states are highly stringy.

With the singularity regulated at the string scale\ $L_c\sim l_s$, 
we see that no string oscillators are turned on in the regulated region, 
\ie\ $n\sim 1$ is already not a light state, from (\ref{lightosc}). If 
instead the regulator is the Planck length\ $L_c\sim 
l_p$,\ then the oscillator state of highest level turned on is\ 
$n\sim ({l_s\over l_p})^2\sim {1\over g_s^{2/(D-2)}}$, using the naive 
relation for the Newton constant\ $G_D=l_p^{D-2}=g_s^2 l_s^{D-2}$. This 
implicitly requires $p_-\ll M_p$. Thus in the weakly coupled (or 
free) string limit\ $g_s\ra 0$, we have $n\gg 1$ in the regulated 
region with a large number of highly stringy oscillator states 
excited.

In a reduced quantum mechanics of the oscillator modes, the wave function 
of the $n$-th highly stringy oscillator state again has an oscillating 
phase (using sec.~3.3 of \cite{knNullws}) but one that is non-divergent 
now, due to the regulator. The overall damping of the real Gaussian part 
is also finite. 
We recall from \cite{knNullws} that the wave functions for the low 
lying oscillator states are regular even in spacetimes with 
singularities for $2\nu\geq 0$. In the regulated spacetime, the 
wavefunctions are also well-behaved for the cases $2\nu<0$.

\section{Discussion}

We have described the Schrodinger picture wavefunctional for string
propagation across null singularities, reconciling this with the
description in terms of the conventional Hamiltonian of oscillator
states. The nonsingular behaviour of the wavefunctional suggests that
free string propagation is well-defined across null cosmological
singularities satisfying certain relations among their Kasner
exponents ($2\nu\geq 0$). These are in fact the singularities for
which the classical near-singularity string mode functions are
non-divergent. While most of our discussion has been for two scale
factors (or Kasner exponents), it is straightforward to generalize
this to multiple Kasner exponents. In other variables, such as those
arising in Brinkman coordinates, the presence of a ``wildly''
oscillating phase makes such an interpretation of the Schrodinger
wavefunctional difficult. We then discussed the role of length scales
and also studied string propagation in spacetimes of this sort where
the singularity has been regulated at some scale (with a certain
regulator).  This gives a slightly clearer picture of how string
oscillator modes get excited near the singularity. In particular for a
finite substringy regulator, there is a finite (but large) number of
string oscillator states excited near the singularity. If the
regularization occurs at the Planck scale, the highest such oscillator
state turned on is of level\ $n\sim ({l_s\over l_p})^2$ . Thus
although the Schrodinger wavefunctional suggests well-defined free
string propagation for some of these singularities, it is conceivable
that the total production of light string states is divergent, as
already suggested in \cite{horowitzsteif}. An important related issue
involves understanding the backreaction on the background of such
light string modes. Also note that our discussion of the Schrodinger
wavefunctional is essentially at the level of first quantized single
strings. The fact that there is a proliferation of light string
oscillator states suggests that perhaps a second quantized framework,
\eg\ string field theory, incorporating string interactions, might be
useful for a more complete understanding of string propagation across
such singularities. A simple, if trite, possibility is simply that
strings get highly excited in the near singularity region but pass
through without significant interaction, then smoothly get de-excited
as we evolve past the singularity towards late times.
It would be interesting to explore these further, and also understand 
the apparently ill-defined singularities (with $2\nu<0$).

We have essentially used the scale factors $h_m(x^+)$ in our solutions
here to simulate the role of the dilaton there in that the internal
$h_m(x^+)$ scale factors shrinking effectively drive the singularity
in the $x^i$-directions, just as the time-varying dilaton drives the
singularity in the AdS/CFT cosmological context
\cite{dmnt,adnt,adnnt}.  Based on our discussion here, it would seem
that the near singularity region in the bulk null AdS cosmologies
\cite{dmnt}, having a sufficiently high local energy density, is
filled with (relatively light) string oscillator states.  Assuming
that the bulk string theory has no qualitatively new features stemming
from the worldsheet coupling to the D3-brane 5-form flux, this system
might be qualitatively similar to the present case.  It would then
seem that interaction effects between the various string modes could
become non-negligible near the null singularity in the bulk. In these
cases however, the string (gauge) coupling\
$g_s(x^+)=e^{\Phi(x^+)}=g_{YM}^2(x^+)$ vanishes near the singularity
so that bulk string interactions might be suppressed, modulo bulk
reflections of possible gauge theory renormalization effects as
discussed (for AdS cosmologies with spacelike singularities) in
\cite{adnnt}. Analogs of such solutions here with null Kasner-like
dilatonic cosmological singularities satisfy\
$R_{++}={1\over 2}(\del_+\Phi)^2$ . With\ $e^\Phi= g_s (x^+)^\al$, 
the Kasner exponents $a,b$, now satisfy\ 
$a^2+2a+{D-4\over 2}(-b^2+2b+2ab) = {\al^2\over 2}$, and similarly 
for multiple Kasner exponent solutions.

In general, one expects\footnote{I thank A.~Sen for discussions on
  this point.} that null singularities have no $\al'$ corrections: the
lightlike nature forbids any nonzero covariant contraction
contributing to a higher derivative correction to the effective action
that might correct the singular region of the spacetime.  As long as
string interaction ($g_s$) effects also give only local covariant
corrections to the low energy effective action, these will also
vanish. This might lead us to imagine that null singularities are
perhaps not resolved at all since possible stringy corrections vanish,
\ie\ these are not allowed singularities in string theory. In this
sense, these are quite different from spacelike singularities where
higher derivative corrections (\ie\ stringy effects) yield 
increasingly important corrections near the singularity. 
This argument relies on the presence of the null isometry: however 
the low energy null isometry could be invalid in the near singularity
region, \eg\ broken by stringy effects.

From the worldsheet analysis above (and in \cite{knNullws}), we see
detailed distinctions between the behaviour of string modes depending
on the Kasner exponents leading us to suspect that such a general
no-go argument need not be strictly true. Most notably, since string
oscillator modes are being excited, in particular with a set of highly
stringy modes being light, it is conceivable that nonlocal stringy
effects become important near the singularity. The rough intuition is
that a string mode of high oscillation number corresponds to a highly
extended (or long and wiggly) string: such highly extended strings
would in general be expected to intersect and thus interact
nontrivially, fitting naturally within a second quantized framework
incorporating interactions.  This would be consistent with the idea
that the low energy notion of spacetime does not exist in the vicinity
of the singularity. Thus the possibility of a low energy mechanism for
null singularity resolution via higher derivative corrections to a low
energy effective description need not exist either. Similar features
can of course be recalled from investigations of \eg\ flop transitions
in Calabi-Yau spaces.


\newpage
\noindent {\small {\bf Acknowledgments:} It is a great pleasure to 
thank S. Das and especially S. Trivedi for several very useful 
discussions. I have also benefitted from conversations with B. Craps, 
G. Mandal, L. McAllister, S. Minwalla and A. Sen. I'd like to
thank the hospitality of TIFR, Mumbai, as this paper was being 
finalized. This work is partially supported by a Ramanujan Fellowship, 
DST, Govt of India.}

\end{document}